\documentstyle[aps,prl,epsf,preprint]{revtex}
\def\fun#1#2{\lower3.6pt\vbox{\baselineskip0pt\lineskip.9pt
        \ialign{$\mathsurround=0pt#1\hfill##\hfil$\crcr#2\crcr\sim\crcr}}}
\begin{document}
\title{\vskip-2.5truecm{\hfill \baselineskip 14pt {{
\small  FERMILAB-Pub-97/125-A\\
       \hfill CERN-TH/97-130}}\vskip .1truecm} 
\vskip 0.1truecm {\bf Minimalism in Inflation Model Building \thanks{Submitted to Phys. Lett. {\bf B}.}}}
\author{{Gia Dvali}\thanks{Georgi.Dvali@cern.ch.}$^{(1)}$ and 
{Antonio Riotto}\thanks{
 riotto@fnas01.fnal.gov.}$^{(2)}$}
\address{$^{(1)}${\it Theory Division, CERN\\Geneva, Switzerland}}
\address{$^{(2)}${\it NASA/Fermilab Astrophysics Center, Fermilab 
National Accelerator Laboratory,\\Batavia, IL 60510, USA}}
\maketitle


\begin{abstract}
\baselineskip 12pt

In this paper we demand that a successfull  inflationary scenario should
follow from a model entirely motivated by particle physics considerations.
We show  that such a connection is indeed possible within the
framework of concrete supersymmetric Grand Unified Theories where  the  doublet-triplet splitting problem is naturally solved.
The Fayet-Iliopoulos $D$-term  of a gauge
$U(1)_{\xi}$ symmetry, which  plays a crucial role in the solution of the  doublet-triplet splitting problem,   
simultaneously provides a built-in inflationary slope  protected from 
dangerous supergravity corrections.

\end{abstract}
\thispagestyle{empty}
\newpage
\pagestyle{plain}
\setcounter{page}{1}
\def\beq{\begin{equation}}
\def\eeq{\end{equation}}
\def\beqa{\begin{eqnarray}}
\def\eeqa{\end{eqnarray}}
\def\tr{{\rm tr}}
\def\x{{\bf x}}
\def\p{{\bf p}}
\def\k{{\bf k}}
\def\z{{\bf z}}
\baselineskip 20pt
\section{Introduction}

The flatness and the horizon problems of the standard big  bang cosmology 
are elegantly solved if during the evolution of the early Universe the energy
density happens to be dominated  by some  vacuum energy and comoving scales grow quasi-exponentially  \cite{guth81}. An  inflationary stage 
is also required to diluite any undesirable topological defects left as remnants after
some phase transition taking place at early epochs.  The vacuum energy driving inflation is generally assumed to be associated to 
some scalar field, the inflaton,  which is  displaced from the minimum of its 
potential. As a by-product, quantum fluctuations of the inflaton field  may be the seeds  for the generation of structure
formation.

There are many problems one has to face in building up a successfull inflationary model. First of all, the level of density and temperature fluctuations observed
in the present Universe, $\delta\rho/\rho\sim 10^{-5}$, require the inflaton
potential to be extremely flat. This is in contrast with the requirement that the couplings of the inflaton field to other degrees of freedom cannot be too  small otherwise the reheating process,  which converts the vacuum energy into radiation at the end
 of inflation,  takes place too slowly: large 
couplings induce large  loop corrections to the inflaton
potential,  spoiling its flatness. Introducing very small
parameters to ensure the extreme flatness of the inflaton potential seems very
unnatural and fine-tuned in  most non-supersymmetric theories. However,  this technical
naturalness may be achieved in supersymmetric models \cite{ellis82} because the
nonrenormalization theorem guarantees that the superpotential is not
renormalized to all orders of perturbation theory \cite{grisaru79}. 
The perturbative renormalization of the K\"ahler potential, however, can
be crucial for the inflationary dynamics due to a non-zero energy
density which breaks supersymmetry spontaneously during inflation \cite{dss,ddr}\footnote{In particular, this renormalization can be due
to same strongly coupled particles whose condensate generates the inflaton
scale dynamically \cite{ddr}.},
independently whether this energy density is an input or
results from some strong dynamics 
\cite{dsi,ddr}. Secondly,
there is the (aesthetic) problem of   embedding   a successfull inflationary scenario in the framework of some  well-motivated   particle physics models.  

To our opinion, one should apply a sort of  "minimal principle" \cite{dr} 
requiring that any successfull inflationary scenario should 
naturally arise from models which are  entirely
motivated by particle physics considerations and  should not   involve (usually complicated and {\it ad hoc}) sectors on top of the existing structures.
 Recently such attempts have been made in \cite{ddr},
in the framework of dynamical grand unified symmetry breaking, and in \cite{dr}
where the inflaton candidates were identified in some models of gauge-mediated supersymmetry breaking.

It is the main purpose of this paper to demonstrate a possibility of the
connection between the inflationary scenario and the particle physics problems, within the framework of concrete Grand Unified Theories. In doing that, we will be entirely motivated
by the solution to a serious problem arising in supersymmetric Grand Unified Theories (SUSY GUTs),  namely the doublet-triplet splitting problem. We will show that
the model which is able to solve this problem also naturally 
 incorporates a built-in inflationary scenario. We will also show that our proposal  escapes
the usual slow-roll problems posed by supergravity corrections in $F$-term dominated inflation. The supergravity corrections usually  induce  large (of order the Hubble parameter $H$)
curvature for the inflaton slope and inflation does not take place \cite{Hproblem}. The appearance of such a large curvature reflects the fact that SUSY must be broken during inflation. This mass does not disappear in the limit in which the Planck mass 
$M_{P\ell}$ tends to infinity when $H$ is held  fixed. As was suggested in \cite{bd}, 
one possible way out  to avoid this  problem is to have inflation
 dominated by a $D$-term \footnote{See also comment in \cite{st}.}. Indeed, in the de Sitter space the gravity-transmitted
$D$-type supersymmetry breaking can be much weaker than the $F$-type counterpart and 
 the slow-roll problem may be avoided. Large $D$-term driving inflation  can be induced,
for example,  if the theory contains a gauge $U(1)_{\xi}$ factor with
a nonvanishing  Fayet-Iliopoulos $D$-term 
\begin{equation}
\int d^4\theta \xi V.  \label{fit}
\end{equation}
This term may be present in the underlying theory from the very beginning (
it is allowed by a gauge symmetry, unless $U(1)$ is embedded in some non-Abelian group\footnote{$\xi = 0$ can be enforced by charge conjugation symmetry \cite{witten}
which flippes all $U(1)$ charges. Such symmetry is possible in nonchiral theories.}) or may appear in the effective theory after some heavy degrees of freedom have been integrated out. Moreover,  
it looks  particularly intriguing
that an  anomalous $U(1)_\xi$
symmetry  is  usually present in string theories \cite{fi}\footnote{Some
cosmological implications of the anomalous $U(1)$
were studied in a different context \cite{Acos}.}
 and the anomaly
cancelation is due to the Green-Schwarz mechanism \cite{u(1)A} .   The corresponding Fayet-Iliopoulos term is given by
\begin{equation}
\xi = \frac{g^2}{192\pi^2}\:{\rm Tr} {\bf Q}\:M^2, \label{stringfit} 
\end{equation}
where $M=M_{P\ell}/\sqrt{8\pi}$ is the reduced Planck mass and  ${\rm Tr} {\bf Q}\neq 0$ indicates the trace over the $U(1)_{\xi}$ charges of the fields present in the spectrum of the theory. 

On the other hand, the  
  anomalous $U(1)_\xi$ can play also a crucial role in  the solution of
the doublet-triplet splitting problem \cite{solution}.
It is therefore   natural to  attempt to reconcile these two implications coming from  theories containing an anomalous $U(1)_\xi$ symmetry  and to 
construct a model that would solve the hierarchy problem and  simultaneously predict a successfull stage of inflation in the early universe. Before proceeding, we would like to 
to point out that  in our scenario the use of an  anomalous $U(1)_\xi$ is not strictly necessary. What is really crucial  is the  presence of a gauge $U(1)_\xi$ with nonvanishing 
 $D$-term (\ref{fit}).
In this respect any gauge $U(1)_\xi$ would be suitable for our purposes, but  the advantage of an anomalous $U(1)_\xi$ is that
$\xi$ is not an input parameter but is fixed from the expression (\ref{stringfit}). Therefore,
we keep our discussion as general as possible   and explicitly indicate the difference between an anomalous and nonanomalous $U(1)_\xi$ when  the difference is important.

Our inflationary scenario can be regarded as a realistic variant of 
hybrid inflation \cite{linde}. Typically in this scenario the inflaton field
is represented by a gauge singlet coupled to the Higgs field that 
triggers the end of inflation via a  {\it non-thermal} phase transition with symmetry breaking. Dangerous topological defects ({\it e.g.} magnetic
monopoles in the grand unified context) may be  produced.
What is unusual in our scenario is that inflaton is not a gauge singlet, but
resides in the component of the adjoint Higgs that breaks GUT symmetry.
Consequently the GUT symmetry is broken both during and after inflation,
and no monopoles are produced.

\section{The Model}

\subsection{Higgs Sector and the Doublet-Triplet Splitting}

Let us briefly describe the main features of the model we have in mind to solve the doublet-triplet splitting problem. It is essentially based on the mechanism of \cite{custodial}. The novelty in our case is that we incorporate $D$-term in the spirit of \
cite{solution} in order to generate VEVs and therefore simplify the structure of the superpotential.

 Let us consider an $SU(6)$ supersymmetric GUT with one adjoint Higgs
$\Sigma$ and a number of fundamental Higgses 
$H_A, \bar H^A, H_A', \bar H^{A'}$. We assume that each of these
fundamentals transforms as a doublet of a certain custodial $SU(2)_c$
symmetry that is required to solve the hierarchy problem \cite{custodial}. The index 
$A =  1,2$ is the   $SU(2)_c$-index.
We also assume that $H_A, \bar H^A$ carry unit charges
opposite to $\xi$ and are the ones that compensate $U(1)_{\xi}$ $D$-term
in the present Universe. In the context of string inspired anomalous $U(1)_\xi$
this would simply mean that they carry charges opposite to total trace 
${\rm Tr}{\bf  Q}$.

The superpotential reads
\begin{equation}
 W = c{\rm Tr}\Sigma^3 + (\alpha\Sigma + aX + M)H_A\bar H^{'A}
+ (\alpha'\Sigma + a'X + M')H_A'\bar H^A. 
\end{equation}
Minimizing both the $D$- and the  $F$-terms we get the following supersymmetric
vacuum which leaves  $SU(3)_c\otimes SU(2)_L\otimes U(1)$ as unbroken gauge symmetry
\begin{eqnarray}
H_{Ai} &=&\bar H^{Ai} = \delta_{A1} \delta_{i1}\sqrt{{\xi \over 2}},~~~~~
H_A' = \bar H^{A'} = 0,~~~\nonumber\\
\Sigma &=& {aM' - a'M \over a'\alpha - \alpha' a} {\rm diag}(1,1,1,-1,-1,-1),~~~~~
X = -{\alpha M' - \alpha 'M \over a'\alpha - \alpha' a}. \label{vac}
\end{eqnarray}
Here $i,k = 1,2,..6$ are $SU(6)$ indexes.
 The role of the $\Sigma$ VEV is crucial since it leaves
the unbroken $SU(3)_c\otimes SU(3)_L\otimes U(1)_Y$ symmetry,
consequently it can cancel masses of all upper three or lower three
components of the fundamentals\cite{savas}.
The fundamental VEVs are $SU(5)$ symmetric, so that the
intersection gives the  unbroken standard model symmetry group.

In this vacuum the electroweak Higgs doublets from 
$H_2,\bar H^2, H_2^{'}, \bar H^{'2}$
are massless. This is an effect of custodial $SU(2)_c$
symmetry. Indeed, since $H_1$ and $\bar H^1$ break  one of the
$SU(3)$ subgroups to $SU(2)_L$, their electroweak doublet components
become eaten up Goldstone multiplets and cannot get masses from the
superpotential due to the Goldstone theorem. This forces the VEVs of
$\Sigma$ and $X$ to exactly cancel their mass terms and those of $H_2,\bar H^2, H_2^{'}, \bar H^{'2}$ due to the  custodial
symmetry. This solves the doublet-triplet splitting problem
in a natural way.

An alternative possibility would be to relax the requirement of $SU(2)_c$
custodial symmetry and instead to introduce a number of singlets
$X_A, X_A'$ coupled to the different pairs, as it was suggested by Barr
\cite{barr}. In this case one has to assume a nonzero VEV for all
$H_A,\bar{H}^A$ fields. Then the doublet masses will be cancelled
by singlets just as in our example.

\subsection{Fermion Masses}

 Quarks and leptons of each generation are placed in a minimal anomaly
free set of $SU(6)$ group: $15$-plet plus two $\bar 6_A$-plets per family.
We assume that $\bar 6_A$ form a  doublet under $SU(2)_c$ so that
$A = 1, 2$ is identified as $SU(2)_c$ index \footnote{Note that $15 + \bar 6_A$ just form
a fundamental $27$-plet of $E_6$ if
$SU(6)\otimes SU(2)_c$ is viewed as one of its maximal subgroups.}.
The fermion masses are  then  generated as in ref. \cite{custodial} through the
couplings ($SU(6)$ and family indices are suppressed)
\begin{equation}
  \bar H^A\cdot 15\cdot\bar 6_A  + \epsilon^{AB}{H_A\cdot H_B \over M_{\xi}} 15\cdot 15,
\end{equation}
where $M_{\xi}$ has to be  understood as the mass of order $\sqrt{\xi}$ of integrated-out
heavy states (the simplest possibility is to use the  $20$-plet transforming as
doublet under custodial $SU(2)_c$). In the case of anomalous $U(1)_{\xi}$ the relative charges of the
matter field must be fixed from the Green-Schwarz anomaly cancelation. 
When the large VEVs of $H_1$ and $\bar H^1$ are inserted, the additional,
vectorlike under $SU(5)$-subgroup, states: $5$-s from $15$-s and $\bar 5$-s
from $\bar 6_1$, become heavy and decouple. Low energy couplings are just the usual $SU(5)$-invariant Yukawa interactions of the light doublets from $H_2$ and $\bar H^2$ with the usual quarks and leptons.

\section{$D$-term Driven Inflation}

Let us now  show that model briefly described in the previous section  has a built-in inflationary trajectory
in the field space along which all $F$-terms are vanishing and
only the associated $U(1)_{\xi}$ $D$-term is nonzero. As said in the introduction, this peculiar feature will allow inflation to take place   without  suffering from the slow-roll problem induced by the supergravity
corrections.

The relevant branch in the field space is represented by the $SU(6)$ $D$- 
and $F$-flat trajectory parameterized by the invariant ${\rm Tr}\Sigma^2$. This
corresponds to an arbitrary expectation value along the component
\begin{equation}
\Sigma = {\rm diag} (1,1,1,-1,-1,-1) \frac{S}{\sqrt{6}}.
\end{equation}
The key point here is that above component has no self-interaction ({\it i.e.} ${\rm Tr}\Sigma^3 = 0$)
and appears in the superpotential linearly. At the generic point of this
moduli space the gauge $SU(6)$ symmetry is broken to
$SU(3)\otimes SU(3)\otimes U(1)$. All gauge-non
singlet Higgs fields are getting masses ${\cal O}(S)$  and
therefore, for large values of $S$,  $S \gg \sqrt{\xi}$, they decouple. Part of them gets eaten
up
by the massive gauge superfields. These are the  components of $\Sigma$
transforming as $(3,\bar 3)$ and $(\bar 3, 3)$ under the unbroken subgroup. All
other Higgs fields get large masses from the
superpotential. The massless degrees of freedom
along the branch are therefore :
two singlets $S$ and $X$, the massless $SU(3)\otimes SU(3)\otimes U(1)$ super- Yang-Mills
multiplet and the massless matter superfields.

By integrating out the heavy superfields,  we can write down an effective
low energy superpotential by simply 
using holomorphy and symmetry arguments\cite{seiberg}.  This superpotential,
as well as all gauge $SU(6)$ $D$-terms,
is vanishing.  Were not for the $U(1)_{\xi}$-gauge symmetry, the 
branch parameterized by $S$, would simply correspond to a SUSY preserving flat
vacuum direction remaining flat to all orders in perturbation theory.
The $D$-term, however, lifts this flat direction, taking an asymptotically constant value
for arbitrarily large $S$ at the tree-level. This is because all Higgs
fields with   charges opposite to  $\xi$ gain large masses and decouple,
and $\xi$ can not be compensated any more
(notice that  heavy fields decouple in pairs with opposite charges and therefore 
${\rm Tr} {\bf  Q}$ over the remaining low energy fields is not changed).
 As a result, the branch of interest is represented by two massless
degrees of freedom $X$ and $S$ whose VEVs  set the mass scale for the heavy
particles, and a constant tree level vacuum energy density
\begin{equation}
V_{{\rm tree}} = {g^2 \over 2}\langle D ^2\rangle = \frac{g^2}{2}\xi^2.
\label{dpot}
\end{equation}
This term is responsible for inflation.

The above result,  which was based on holomorphy and symmetry arguments, can be easily
rederived by explicit solution of the equations of motion along the inflationary
branch. For doing this,  we can explicitly minimize all $D$- and $F$- terms subject to
large values of $S$ and $X$.  The relevant part of the potential is
\begin{equation}
V = |F_{H_A^{'}}|^2 +  |F_{\bar H_A^{'}}|^2  +  {g^2 \over 2}D^2, \label{fandd} 
\end{equation}
since the remaining $F$- and $D$- terms are automatically vanishing as long as
all other gauge-non singlet Higgses are zero.
We would need to include them only if the  minima of the potential  (\ref{fandd}) (subject to $S,X \gg
\xi$) were
incompatible with such an assumption. However for the branch of our interest
this turns out to be not  the case.

 It is easy now  to check that for
\begin{equation}
{\rm Min} \left ( \left|M + aX  \pm \alpha{S \over \sqrt{6}}\right|, \left| M' + a'X  \pm  \alpha'
{S \over \sqrt{6}}\right| \right ) > g\sqrt{\xi} \label{cri}
\end{equation}
all other VEVs vanish and, therefore, a nonzero contribution to the
potential comes purely from the constant $U(1)_{\xi}$
$D$-term. This is when inflation takes place: starting from some chaotic initial values of
$S$ and $X$ for which the condition (\ref{cri})
is far from being satisfying, the
system
will slowly evolve and inflate. In each case the inflaton field is represented
by the appropriate combination of $S$ and $X$ fields.

Whenever the condition
(\ref{cri}) is violated, some of the $H, \bar H$ components become tachionic and
compensate the $D$-term. The  system  very rapidly relaxes to the supersymmetric
vacuum (\ref{vac}) and oscillates about it. Inflation is therefore terminated by this rapid water-fall \cite{linde} and the universe undergoes a short period of reheating after which it is filled up by particles in thermal equilibrium. 

As we have seen,  the tree-level  potential along the inflationary branch is
exactly flat. Radiative corrections\cite{dss,bd}, however, create a logarithmic slope
that drives inflaton toward the minimum (\ref{vac}).
The origin of this correction can be understood in the following way. As we have shown,
the $S$ and $X$ VEVs set the mass scale for the heavy particles along the
inflationary branch. Thus,  we can think of the low energy theories at the different
points of this branch as of the same theory at the different energy
scales. The gauge coupling in (\ref{dpot}) should be understood as the running
gauge
coupling. This is simply  due to the gauge field wave function renormalization
through the loops with $U(1)_{\xi}$ -
charged particles $H,\bar H, H', \bar H'$.
Since their mass is set by $S$ and $X$ VEVs, the nontrivial dependence on
these VEVs arises, providing effective one-loop potential for the
inflaton field. For large field strengths or, in other words, masses of the particles in the loop
much larger than  $\sqrt{\xi}$,  this potential assumes the following form
\cite{dss},\cite{bd}
\begin{equation}
\label{pot}
 V_{{\rm inf}} = {g^2 \over 2}\xi^2\left( 1 + {3\:g^2 \over \pi^2}
{\rm ln}\left (|\pm\alpha S/\sqrt{6}
+ aX + M||\pm\alpha S/\sqrt{6} + a'X + M'|\right)\right )
\end{equation}
This  is  simply the asymptotic  form for $S,X\gg\sqrt{\xi}$ of the one-loop
corrected effective potential
\begin{equation}
V_{{\rm one-loop}} = {\rm Tr}\:(-1)^F\:{\cal M}^4\: {\rm ln}\:{\cal M}^2. \label{ol}
\end{equation}
The contribution to (\ref{ol}) comes purely from the $H,\bar H', H',\bar H$ 
superfields. These are the  fragments $(1,3),(1,\bar 3)$ and $(3,1),(\bar 3, 1)$
of the  $H, \bar H'$ with supersymmetric masses
\begin{equation}
 \pm\alpha S/\sqrt{6}
+ aX + M,
\end{equation}
and the analogous fragments of the  $H', \bar H$ with supersymmetric  masses
\begin{equation}
\pm\alpha' S/\sqrt{6} + a'X + M',
\end{equation}
respectively.
All these superfields suffer from the tree level non-supersymmetric contribution
to the scalar masses from the $U(1)_{\xi}$  $D$-term equal to
\begin{equation}
\pm g^2\xi,
\end{equation}
where the sign corresponds to the $U(1)_{\xi}$-charge. All other states either have
no mass-splitting due to a vanishing charge (these are $X, \Sigma$ and
the gauge fields) or have no inflaton dependent mass (these are matter fields).

 As we have seen,  the tree level inflationary branch is a two dimensional complex plane subject to the constraint $S,X\gg\sqrt{\xi}$. Classically, any
path parameterized by an arbitrary combination of $S$ and $X$ on this manifold
is exactly flat and can lead to inflation with a nearly equal chance.
So classically inflation can end only when condition (\ref{cri}) breaks down,
signaling that some of the fields become tachionic and system relaxes to the
global minimum. However, as we have argued,  the quantum corrections provide
a slope for the inflaton field and inflation in reality may end much before
the instability occurs,  simply because of the  breakdown of the slow roll
conditions.

Let us denote the direction along which inflation is taking place by $\phi$ and write symbolically the potential (\ref{pot}) as $V_{{\rm inf}} \simeq V_0(1+c\:g^2{\rm log}\phi)$, where $c={6 \over \pi^2}$. 
During the slow-roll phase, when the inflaton  is rolling down from large values, the cosmic scale factor may grow by $N$ e-foldings:  
\begin{equation}
\label{n}
N\simeq\frac{8\pi}{M_{P\ell}^2}\int_{\phi_N}^{\phi_e}\:\frac{V_0}{V^\prime}=
\frac{4\pi}{M_{P\ell}^2}\frac{\phi_N^2}{g^2c},
\end{equation}
where $\phi_e$ denotes the value of the field when inflation ends. Successful inflation requires $N\simeq 60$. 

Fluctuations arise due to quantum fluctuations in the inflaton field. 
We may  then  compute the power spectrum of quantum fluctuations, which is the Fourier transform of the two-point density autocorrelation function. It has the primordial form $P(k)\propto k^n$. where $k$ is the amplitude of the Fourier wavevector and $n$ 

denotes the spectral index.
 The measurement of the quadrupole anisotropy in the cosmic
microwave background radiation detected by COBE \cite{cobe} allows us to fix the parameters of the model:
\begin{eqnarray}
\left(\frac{\Delta T}{T}\right)&=&\sqrt{\frac{32\pi}{45}}\frac{V_0^{3/2}}
{V^\prime(\phi_N) M_{P\ell}^3}\nonumber\\
&\simeq & 0.3 \sqrt{\frac{N}{c}}\left(\frac{\xi}{M_{P\ell}^2}\right).
\end{eqnarray}
Imposing $\left(\frac{\Delta T}{T}\right)\simeq 6\times 10^{-6}$,
for $c\sim \frac {6}{\pi^2}$ 
we get $\sqrt{\xi}\sim 10^{16}$ GeV, which is close to the GUT scale. The spectral index is practically indistinguishable from unity, $n-1\simeq 1-\frac{1}{N}\simeq 0.98$. This recovers prediction of the scenario
\cite{dss}, the difference is that, since our inflation is $D$-dominated,
we do not need any assumption about the non-minimal (quartic) terms in the K\"ahler potential $(\phi^*\phi)^2$.
They do not contribute in the curvature, since $F_{\phi}$ is vanishing
during inflation. On the contrary,  in the $F$-dominated scenario \cite{dss}
the predictions are sensitive to the precise structure of this term
\cite{lindetoni}.

One may ask whether the  value of $\sqrt{\xi}$ required by density perturbations can be motivated by realistic string theory.
At this point uncertainties come from the fact that in our approximation
we were treating $\xi$ as constant (up to a course-graining scale dependence through the gauge-coupling). This is certainly justified in the effective
field theory approach in which $\xi$ is treated as an input parameter.
In string theories the
gauge and gravitational coupling constants are set through the expectation value of the dilaton field $s$ and the Fayet-Iliopoulos $D$-term actually is a function of $(s+\overline{s})$.  Since the dilaton potential most likely is strongly influenced 
by the inflationary dynamics, the  actual value of $\xi$ at the moment
 when observationally interesting scales crossed the horizon during inflation  might be quite different from the  one "observed" today. 
It seems that  entire question is related to  the problem  of the  dilaton
stabilization and it is hard to make any definite statement without knowing the details of the dilaton dynamics during inflation. All our estimates made above are valid
within an effective field
theory description, in which the gauge and gravitational constants can be treated
as parameters whose inflaton-dependence arises from the course-graining
scale-dependence.

\section{The Monopole Problem}

 In the usual hybrid inflationary scenarios \cite{linde} inflation is terminated by the rolling down of a Higgs field coupled to the inflaton and consequent  phase transition with symmetry
breaking. Whenever the vacuum
manifold has a non-trivial homotopy, the topological defects will form
much in the same way as in the conventional thermal phase transition. Thus,  the straightforward generalization of the hybrid
scenario in the GUT context would result in the post-inflationary
formation of the unwanted magnetic monopoles. In our scenario this disaster never happens, since the inflaton field is the GUT Higgs itself.
The GUT symmetry is broken both during and after inflation and the monopoles
(even if present at the early stages) get inevitably inflated away.
The unbroken symmetry group along the inflationary branch is
$G_{{\rm inf}} = SU(3)\otimes SU(3)\otimes U(1)\otimes SU(2)\otimes U(1)_{\xi}$\footnote{If the gauge $U(1)_{\xi}$ is a stringy anomalous $U(1)$, 
it will be broken by the dilaton even if all other charged fields vanish.
In this case the unbroken symmetry has to be understood as a global one.}
which gets broken to
$G_{{\rm postinf}} = SU(3)\otimes SU(2)\otimes U(1)\otimes U(1)$
modulo the electroweak phase transition (extra $U(1)$ -factor is global).
Since $\pi_2(G_{{\rm inf}}/G_{{\rm postinf}}) = 0$ no monopoles are formed.

In conclusion,  we have shown that a successfull model of
inflation may naturally arise from  concrete
supersymmetric Grand Unified Theories where  the  
oublet-triplet splitting problem is solved.
To achieve that, no price of enlarging the scalar sector 
is to be   paid.  
The Fayet-Iliopoulos $D$-term  of a gauge
$U(1)_{\xi}$ symmetry plays a crucial role both in the solution of the  doublet-triplet splitting problem and 
in providing  a suitable  slope for the inflaton potential which is   protected from 
dangerous supergravity corrections. Since the inflaton is a GUT adjoint
Higgs field, the Grand Unified symmetry is broken both during and
after inflation. As a result, the universe popping out after inflation is safe from  monopoles.

\acknowledgements
AR would like to thank D. Lyth for many enlightening   discussions. AR is  supported  by DOE and NASA grant NAG5-2788 at Fermilab. 

\begin{enumerate}

\end{enumerate}

\end{document}